\documentclass[draft]{agujournal2019}
\usepackage{url} 
\usepackage{lineno}
\usepackage[inline]{trackchanges} 
\usepackage{soul}
\usepackage{amssymb}

\draftfalse

\journalname{JGR: Space Physics}

\begin{document}

%
%

\title{The spacecraft wake: Interference with electric field observations and a possibility to detect cold ions}

%
%




\authors{M. Andr\'e\affil{1}, A. I. Eriksson\affil{1}, Yu. V. Khotyaintsev\affil{1} and S. Toledo-Redondo\affil{2}}

\affiliation{1}{Swedish Institute of Space Physics, Uppsala, Sweden}

\affiliation{2}{Department of Electromagnetism and Electronics, University of Murcia, Murcia, Spain.}




\correspondingauthor{Mats Andr\'e}{mats.andre@irfu.se}




\begin{keypoints}
\item Plasma wakes are common behind scientific spacecraft
\item Wakes in the solar wind can be compensated for in data analysis 
\item Enhanced wakes in the polar lobes can be used to detect cold outflowing ions 
\end{keypoints}

\vspace{50mm}
{\Large An edited version of this paper was published by American Geophysical Union (2021).\\}
\vspace{5mm}
{André, M., A. I. Eriksson, Yu. V. Khotyaintsev and S. Toledo-Redondo (2021) The spacecraft wake: Interference with electric field observations and a possibility to detect cold ions. Journal of Geophysical Research: Space Physics, 126, 2021JA029493, doi: 10.1029/2021JA029493.}

%
%

%
%

\begin{abstract}
Wakes behind spacecraft caused by supersonic drifting positive ions are common in plasmas and disturb in situ measurements.
We review the impact of wakes on observations by the Electric Field and Wave double-probe instruments  on the Cluster satellites. In the solar wind the equivalent spacecraft charging is small compared to the ion drift energy and the wake effects are caused by the spacecraft body and can be compensated for. We present statistics of the  direction, width and electrostatic potential of wakes, and compare with an analytical model. 
In the low-density magnetospheric lobes the equivalent positive spacecraft charging is large compared to the ion drift energy and an enhanced wake forms. In this case observations of the geophysical electric field with the double-probe technique becomes extremely challenging. Rather, the wake can be used to estimate the flux of cold (eV) positive ions. 
For an intermediate range of parameters, when the equivalent charging of the spacecraft is similar to the drift energy of the ions, also the charged wire booms of a double-probe instrument must be taken into account. We discuss an example of these effects from the MMS spacecraft near the magnetopause. 
We find that many observed wake characteristics provide information which can be used for scientific studies.
An important example is the enhanced wakes used to estimate the outflow of ionospheric origin in the magnetospheric lobes to about 10$^{26}$ cold (eV) ions/s, constituting a large fraction of the mass outflow from planet Earth.

\end{abstract}

\section*{Plain Language Summary}
Wakes caused by spacecraft motion or drifting plasma are common behind spacecraft with scientific instruments and disturb in situ observations of space plasmas. 
We review the impact of wakes on observations by the Electric Field and Wave double-probe instruments on the Cluster satellites.
In the solar wind, the wake behind a Cluster spacecraft is caused by the spacecraft body, is narrow, and can partly be compensated for when analysing data. In the regions above the Earth´s polar regions, the wake behind a Cluster spacecraft is caused by an electrostatic structure around the positively charged spacecraft, causing an enhanced wake. The charging stops positive ions from reaching the spacecraft. Rather, this wake can be used to estimate the flux of cold (eV) positive ions escaping from the ionosphere. Above the poles the flux is about 10$^{26}$ ions/s, constituting a large fraction of the mass outflow from planet Earth. 
For an intermediate range of parameters, when the drift energy of the ions is comparable to the equivalent charge of the spacecraft, also the charged wire booms of a double-probe instrument must be taken into account. We discuss an example from the MMS spacecraft near the magnetopause. 

%
%

%


%
%
%
%

\section{Introduction}

Wakes behind obstacles in supersonic flows are common in nature. 
Here we discuss wakes in collisionless plasmas, in particular behind spacecraft. In situ observations are a powerful tool to observe space plasmas, but includes the problem of the spacecraft disturbing the plasma of interest. In many situations spacecraft wakes are caused by flows which are supersonic with respect to the ion thermal speed, but subsonic with respect to the electron thermal speed. The result is that the wake charges negatively until the potential is sufficiently negative to prohibit further accumulation of electrons, hence causing an enhancement of the local electric field. We review the impact of wakes on observations by the Electric Field and Wave (EFW) double-probe instruments using long wire booms in the spin plane of the Cluster satellites. We compare with observations by similar instruments on the MMS satellites. We also briefly discuss the Cluster observations in the context of wakes behind spacecraft in other orbits and behind natural objects.

In some cases wakes are due to the spacecraft body itself and the transverse extent is limited. Here effects on electric field observations can routinely be removed and observations of the geophysical electric field are mainly unaffected \cite{Khotyaintsev2014}. We show examples of Cluster observations in the solar wind \cite{Eriksson2006, Eriksson2007b, Khotyaintsev2010}. 
The direction of the wake gives the direction of solar wind. We show that statistics of the width and electrostatic potential of solar wind wakes are in reasonable agreement with a simple analytical model. While ions in the solar wind often can be studied by particle detectors, we use electric field observations from the solar wind to verify that our understanding of wake physics is correct. 

In other cases the wake is not due to the spacecraft body but to an extended electrostatic structure around a positively charged spacecraft scattering positive ions. Here the wake is extended and observations of the local electric field are complicated to use for investigations of the geophysical E-field. In many situations cold (eV) ions can not reach particle detectors onboard the spaceraft. Rather, the detection of the extended wake can be used to gain information on the cold ions causing the wake. We show examples of Cluster observations in the polar lobes and review how this extended wake can be used for statistical studies of the outflow of cold ionospheric ions \cite{Engwall2009a, Engwall2009b, Andre2015a}. 

In some cases of intermediate parameters, with a positively charged spacecraft but ions that can still reach the satellite, the electrostatic structure around a spacecraft can not be approximated by a sphere but the charged long wire booms of an E-field instrument must be considered. We show an example observed close to the magnetopause by MMS \cite{Toledo-Redondo2019}. For comparison, we briefly discuss wakes in the ionosphere where effects of a negatively charged spacecraft and smaller Debye lengths and gyro radii are important. Overall we find that understanding the physics behind the spacecraft wakes, the local effects on electric field observations can sometimes be removed and most of the observations can be used as originally intended. When this is not possible, sometimes entirely new geophysical parameters such as ion flux can be estimated.

\section{Wakes in different situations}
An object moving in a neutral gas dominated by collisions is either sub- or supersonic. We consider collisionless plasmas. The drift velocity of such a plasma is often larger than the thermal speed of the ions but smaller than the thermal speed of the electrons. Since the drift is supersonic with respect to the ions but subsonic with respect to the electrons, it can be called mesosonic. (We here use the term "supersonic" when comparing ion drift and thermal speeds, since for equal ion and electron temperatures the ion acoustic speed is similar to the ion thermal speed.) A mesosonic drift will cause a negatively charged wake. Hence the presence of a spacecraft in a drifting plasma can cause a local electric field in the vicinity of the spacecraft. 

\subsection{Charged spacecraft}
Spacecraft are usually charged, which affects observations of the local plasma. In Low Earth Orbit in the high density ionosphere, spacecraft are often negatively charged due to the large flux of ionospheric electrons. At higher altitudes in a low density plasma, the photoelectrons emitted by a spacecraft in sunlight can dominate the charging process, causing positive charging. Any deviation from charge neutrality will significantly affect charged particles with an energy similar to the equivalent spacecraft charging. This can in turn influence wake formation and the corresponding local electric field. Spacecraft charging is well known in near-Earth plasmas as discussed below, and also for interplanetary spacecraft such as Rosetta investigating comet 67P \cite{Johansson2020, Bergman2020}.

\subsection{Spacecraft and instruments}
The wakes we consider in detail are related to the ESA Cluster \cite{Escoubet2001} and NASA MMS \cite{Burch2016b} spacecraft, launched 2000 and 2015, respectively. Both are four-spacecraft missions for detailed investigations of space plasma physics. All satellites have long wire booms in the spin plane, used for observations of the electric field \cite{Pedersen1998, Maynard1998}. 
The Cluster Electric Field and Wave (EFW) instrument includes two pairs of probes on wire booms on each satellite. Each pair has a probe-to-probe separation of 88 m, and the electric field is obtained from the potential difference between the probes \cite{Gustafsson1997, Gustafsson2001}. The satellites have a diameter and height of 2.9 and 1.3 m, respectively. The spherical probes have a diameter of 8 cm and the cylindrical  pre-amplifiers located 1.5 m closer to the satellite have the same diameter. To avoid shadow on the probes from the pre-amplifiers, the short stiff booms carrying magnetometers, and from the spacecraft body, the spin plane was initially inclined a few degrees with respect to the ecliptic plane.  Figures~\ref{fig:figure-WakeSketchCombined}a,b show one Cluster satellite in different phases of the $\sim$4-second spin. The MMS spacecraft have a similar diameter, a spin period of $\sim$20 s, and the Spin-plane Double Probe instrument (SDP) has a probe-to-probe separation of 120 m \cite{Lindqvist2016}. The MMS satellites also have an Axial Double Probe instrument with cylindrical sensors separated by 32 m along the spin axis \cite{Ergun2016}. 

Both the Cluster and the MMS spacecraft have additional instruments for observations of quasi-static electric fields, based on a completely different technique. The Electron Drift Instruments (EDI) on Cluster  \cite{Paschmann1997, Paschmann2001} and MMS \cite{Torbert2016} measure the drift of artificially emitted high-energy (0.25--1 keV) electrons as they gyrate back to the spacecraft under the influence of the geophysical magnetic field \cite{Paschmann1998}. These electrons can have gyro radii of several kilometers and are not significantly affected by the local wake. The EDI instruments are therefore not sensitive to spacecraft-plasma interactions but are limited to reasonably steady and strong magnetic fields ($\gtrsim 30$ nT) and quasi-static electric field ($\lesssim 10$ Hz), while double-probe instruments can be used up to MHz frequencies and have additional data products such as spacecraft potential, which can be used for density estimates \cite{Eriksson2006, Pedersen2008}. In addition, both Cluster and MMS have instruments for Active Spacecraft Potential Control (ASPOC), reducing  positive potential by emitting positive ions \cite{Torkar2001, Torkar2016}. 

\subsection{Narrow and enhanced wakes}
Cases of practical importance include spacecraft in the solar wind when the narrow wake is caused by the spacecraft body, and spacecraft in the polar lobes when the wake is caused by an electrostatic structure around a positively charged spacecraft scattering positive ions. These two examples are illustrated in 
Fig.~\ref{fig:figure-WakeSketchCombined}c, d. For simplicity, in this figure we consider the plasma flow to be in the spin plane of the spacecraft. The narrow wake in Fig.~\ref{fig:figure-WakeSketchCombined}c will not affect the electric field observations in the spin phase illustrated in Fig.~\ref{fig:figure-WakeSketchCombined}a when both probe pairs are at a large angle to the flow, but will severely affect observations in the phase shown in Fig.~\ref{fig:figure-WakeSketchCombined}b when one of the probe pairs (3-4) is aligned with the flow. The enhanced wake (Fig.~\ref{fig:figure-WakeSketchCombined}d) will affect the observations for most directions of the wire booms. 

\subsection{Wakes in low Earth orbit}
The basic theory of spacecraft wakes was understood early in the space age \cite{Alpert1965, Gurevich1969} and during the first decades a substantial amount of observations in LEO accumulated \cite{Hastings1995}. Many early wake studies concentrated on these low altitudes since several satellites, including most manned spacecraft, operate in the ionosphere.
At low altitudes in the high density ionosphere a spacecraft typically has negative charge due to the high electron flux, causing the ions to fill the wake more effectively (Fig.~\ref{fig:figure-WakeSketchCombined}e). 
We note that these wakes are very different from the enhanced wakes illustrated in Fig.~\ref{fig:figure-WakeSketchCombined}d.
An orbiting satellite is moving at 7-8 km/s in a rather dense plasma and strong magnetic field, the Debye length and electron gyro radius are typically smaller than the satellite dimensions, while the ion gyro radius can be comparable to the spacecraft dimensions (see Table~\ref{tab:Table_1} for examples of parameters). This is in contrast to the regions at higher altitudes we consider below where Debye lengths and gyro radii are larger than the spacecraft dimensions. 
The small Debye length in LEO gives large wake potentials, which further concentrated early studies to low altitudes. Observations \cite{Katz1998, Ferguson2013} and recent simulations of wakes and related effects include the geomagnetic field for orbiting spacecraft in LEO \cite{Miyake2020}, while other simulation studies consider slower sounding rockets \cite{Darian2017}, and their booms of a few meters \cite{Paulsson2018, Paulsson2019}. Wakes in LEO can also be of practical interest for close-proximity formation flying \cite{Maxwell2019, Maxwell2021}.

\subsection{Wakes behind natural objects}

\begin{figure*}
\begin{center}


\includegraphics[width=0.9\columnwidth]{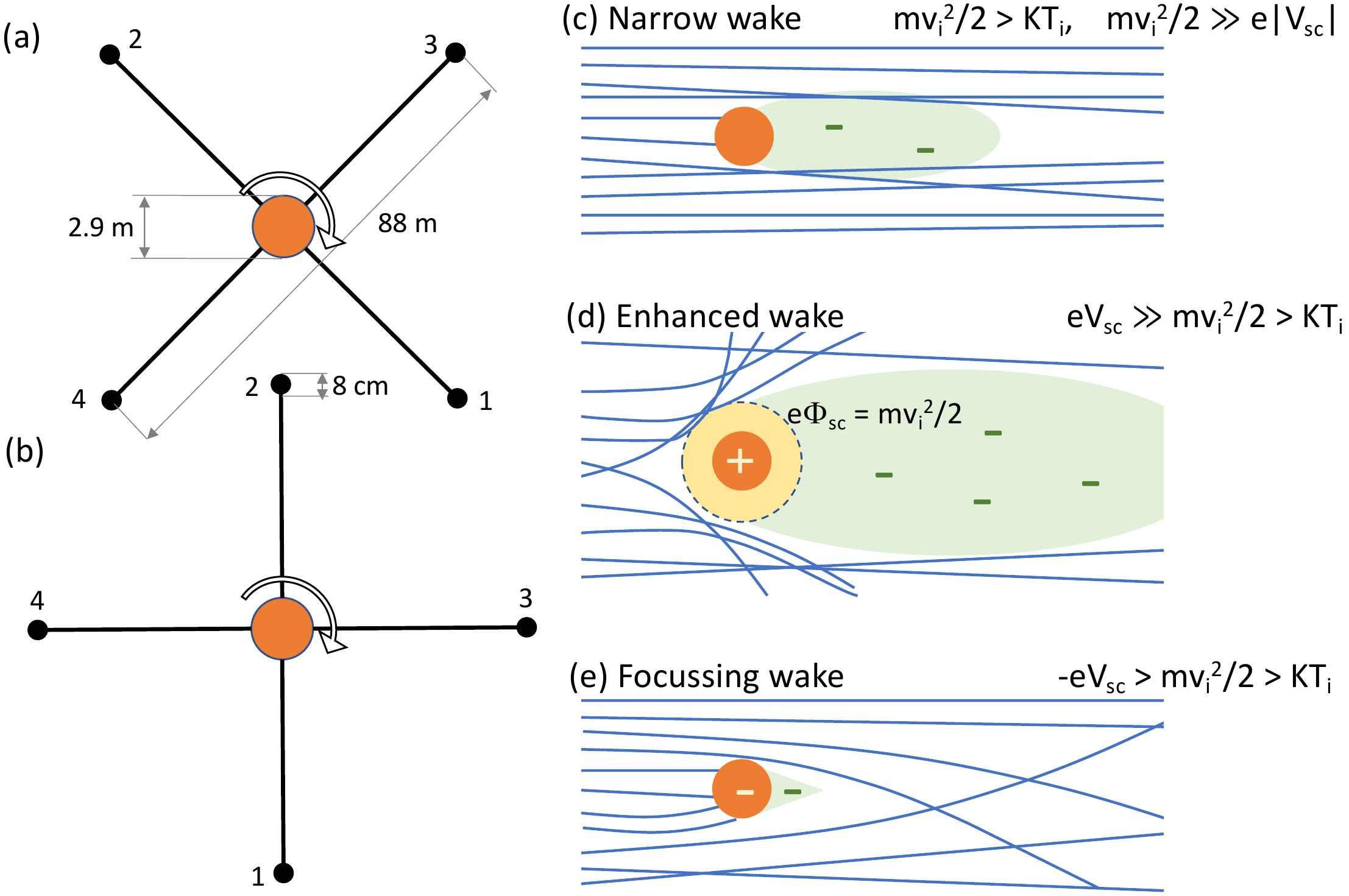}

\end{center}


\caption{Left panel, (a) and (b): Sketch of the Electric Field and Wave instrument on Cluster, using probes on long wire booms, in two different phases of the 4 second satellite spin. 
    Right panel: Some wake cases. Positive ion trajectories are shown in blue, motion is from left to right. The spacecraft is indicated in orange, the green shaded regions indicate negative space charge. (c) When the ion energies are large compared to the equivalent charge of the spacecraft and in the wake, the wake transverse size close to the spacecraft is set by the spacecraft dimensions and the length depends on the ratio of ion flow to thermal speed (e.g. Cluster in the solar wind) (d) For a very positive spacecraft, the ions undergo Rutherford scattering on the potential $\Phi_{sc}$ from the spacecraft, creating an enhanced wake. The dashed circle indicates the equipotential of the spacecraft electrostatic field where $e\Phi_{sc} = mv_i^2/2$ around which ions will scatter (e.g. Cluster in the polar lobes). (e) For the commonly studied ionospheric case, the focusing effect of a negative spacecraft fills in the wake more effectively than in case (c). For all examples the particles are assumed to be unmagnetized which is often a good approximation for wake studies in the solar wind and polar lobes, but not always in the ionosphere. For some parameters also the charging of long wire booms are important for the ion trajectories, see Fig.~\ref{fig:figure-MMS_Wake_v2}}.

    \label{fig:figure-WakeSketchCombined}

\end{figure*}

We concentrate on wakes behind artificial conducting spacecraft and understanding of their effects. This understanding is valuable for interpretation of in situ observations. Overall understanding of wakes is also important for investigations of natural objects not further discussed here. This includes small objects such as charged dust \cite{Miloch2017, Darian2019}. 
This also includes large objects such as the Moon, see \citeA{Rasca2021} and references therein. As another example, investigations of solar wind interactions, including wake formation, with a metal-rich asteroid such as 16 Psyche can be used to understand the present electromagnetic environment and compare scenarios for formation and solidification \cite{Fatemi2018}.
Typically these wakes are very different from the enhanced wakes illustrated in Fig.~\ref{fig:figure-WakeSketchCombined}d and further discussed in sections \ref{sec:Polarwake} and \ref{sec:ColdIons}.

%

\section{Spacecraft wakes in different space plasma}

Polar orbiting spacecraft, such as Cluster, can investigate both the solar wind and the polar lobes. In both cases the density is much lower than in the ionosphere. It was early realized that a spacecraft in a low density plasma generally will be positively charged since satellite photo-emission dominates the influx of electrons from the surrounding plasma \cite{Whipple1965}. However, there are only a few early investigations relevant for wakes behind positive spacecraft potentials, as summarized by \citeA{Engwall2006} and \citeA{Eriksson2007b}. Observations of wakes behind positively charged spacecraft are discussed below. Some relevant simulations of spacecraft wakes and the effects on double probe observations are given by \citeA{Engwall2006}, \citeA{Miyake2013} and \citeA{Miyake2016}. 

\subsection{Wake in the solar wind (narrow wake)} \label{sec:SWwake}

As the solar wind ion flow is supersonic, a wake will form behind a spacecraft in this medium. Because of photoelectron emission, the spacecraft is typically charged to a few volts positive. The ion flow energy $m {v_{i}}^{2}/2$ is usually much larger than the spacecraft-to-plasma potential $e V_{SC}$ (spacecraft charging, where $e$ is the elementary charge) and is also larger than the ion thermal energy  $KT_{i}$ (and the often similar electron thermal energy $KT_{e}$), see Table~\ref{tab:Table_1} for examples of parameters:

\begin{equation}
m {v_i}^{2}/2 \gg eV_{SC}, \hskip 2 cm  m v_i^2/2 >  KT_i \sim KT_e .
\label{eq:Narrow-Wake}
\end{equation}

This supersonic ion drift gives a narrow transverse width of the wake, whose cross-section immediately behind the spacecraft has the size and shape of the spacecraft body, see Fig.~\ref{fig:figure-WakeSketchCombined}c. For typical solar wind speeds and electron temperatures the solar wind is subsonic with respect to the electrons, which therefore can enter the wake. The wake becomes negatively charged.

\begin{figure*}
\begin{center}
	\includegraphics[width=80 mm]{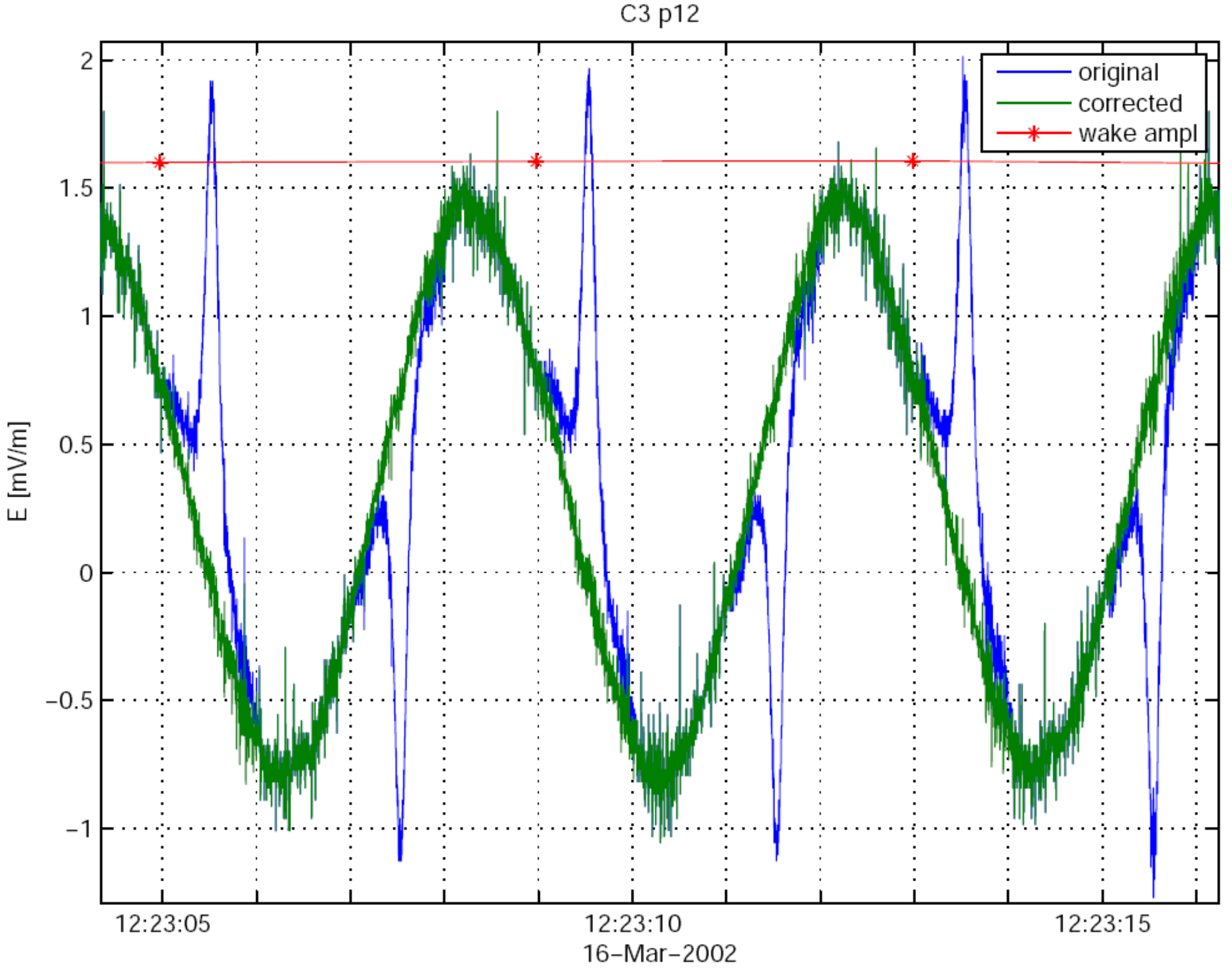}
	\end{center}
    \caption{Solar wind wake signature observed by one probe pair (1-2) of the EFW instrument on one Cluster spacecraft (C3). The blue curve is the original raw data sampled at 450 Hz, while the green curve shows the data after wake removal (see text). The red stars, bounded together with by the red line, shows the wake amplitude determined in the removal process, once for each 4 s spacecraft spin. In the case of a narrow wake, the wake signatures can be compensated for. From \citeA{Eriksson2007b}, see also \citeA{Khotyaintsev2010}.}
    \label{fig:figure-SolarWindWake}
\end{figure*}

The effect on a double probe electric field observation is clear, repetitive at the rate related to the satellite spin period, and easy to identify. Figure~\ref{fig:figure-SolarWindWake} shows an example of wake effects on electric field observations by the Cluster1  EFW probe pair 1-2 in the solar wind. The spikes in the observed electric field (blue) are seen every 2 seconds, or twice per spin period (4 s). This corresponds to each of the probes 1 and 2 encountering the the wake once per spin.   

We have developed an algorithm for the EFW instrument to detect and remove the local wake electric field from the data \cite{Eriksson2006, Khotyaintsev2010, Khotyaintsev2014}. The process involves taking a weighted average of a few 4-second satellite spins which will not affect the very repetitive artificial wake signatures much, while natural wave activity will mainly be removed. Using these averaged data, the artificial signature is identified and then subtracted from the original observation using an algorithm in several steps \cite{Eriksson2007b, Khotyaintsev2010}. The algorithm used to remove the wake from electric field data to be archived collects three primary characteristics of the wake: direction in the satellite spin plane (wake spin angle), amplitude and width (quantified as the full width at half maximum value, FWHM).

The main features of the observed wake can be compared to a simple theoretical model. The ions have a large gyroradius (Table \ref{tab:Table_1}) and as further discussed below the ion trajectories can be well approximated by straight lines on the length-scale of the wire booms. 
In this model a solar wind ion distribution with drift energy $m v_{i}/2$ and thermal energy $KT_{i}$ is stopped by the spacecraft body but no other effect of the spacecraft is included.
Describing the ions by a drifting Maxwellian, the ion density in the wake formed behind the spacecraft can then be calculated by integrating the distribution function over all ion energies and all directions of motion except those blocked by the spacecraft body. Writing the ion density in the wake as $n_i = n_0 - \delta n$ and setting the solar wind to flow in the $+z$ direction, we then have \cite{Alpert1965}
\begin{eqnarray}\label{eq:alpert}
    \delta n(x,y,z) & = & \frac{n_0}{\pi\, z^2} M^2 \exp \left( - M^2\, \frac{x^2+z^2}{z^2} \right) \cdot \nonumber \\ & & \cdot \int_S \exp \left( -M^2\, \frac{x_0^2+z_0^2 -2xx_0-2zz_0}{z^2} \right)\, dx_0\, dy_0
\end{eqnarray}
where $M$ is the ion flow Mach number,
\begin{equation}
    M = \sqrt{\frac{m_i v_i^2}{2 K T_i}},
\end{equation}
and $S$ is the spacecraft cross section in the $xy$ plane. Numerical evaluation of this integral can be used to find the density in the wake at the position of the EFW probes. The ions gradually fill the wake due to their random thermal motion. At the same time the wake widens as ions outside the low density region move into the wake. In this model, the ion charge is not important for the ion motion. In the solar wind this is a good approximation. When reaching potentials $\sim -KT_e/e$ the density of electrons filling the wake reaches an equilibrium. As $KT_e \sim 10$~eV in the solar wind, this negative potential has quite small impact on the motion of the ions with $mv_i^2 /2 \sim 1$~keV.

The quantity measured by EFW is the wake potential $\Phi_{w}$. The electrons are essentially unmagnetized at the scales of interest (Table \ref{tab:Table_1}) and an electron gas in thermal equilibrium is well described by the Boltzmann relation
\begin{equation}\label{eq:boltzmann}
\Phi_{w} = \frac{KT_e}{e}\, \ln 
 \frac{n_e}{n_0} 
\end{equation}
By assuming quasi-neutrality, $n_e \approx n_i$ we can find the wake potential by combining equations~(\ref{eq:alpert}) and (\ref{eq:boltzmann}).
This approximation assumes a short the Debye length, and we return below to how well this last assumption can be expected to hold.

\begin{figure}
    \centering
    \includegraphics[width=13cm]{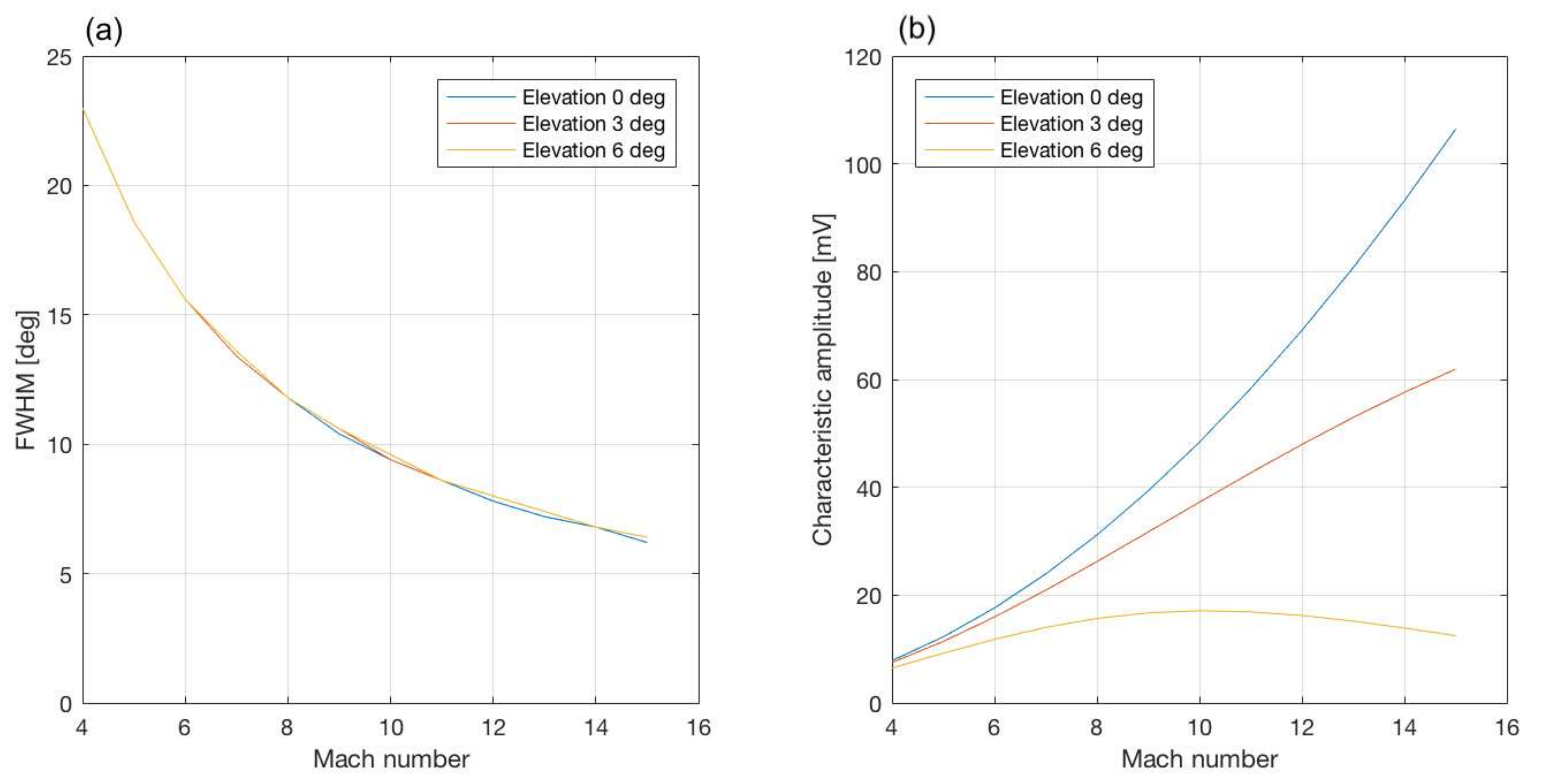}
    \caption{Theoretical wake potential properties at the EFW probes, calculated by numerical integration of Equation~(\ref{eq:alpert}), as function of solar wind ion flow Mach number for three different wake elevation angles of the solar wind direction with respect to the spacecraft spin plane (containing the EFW wire booms). (a) The width (FWHM) of the wake is a robust estimate and lines for all angles are the same within the accuracy of the numerical calculation. (b) The estimated amplitude is a characteristic value relevant for typical solar wind parameters, including $KT_e \approx 10$~eV, not the exact peak potential in the wake.} 
    \label{fig:alpert}
\end{figure}

Predicted EFW observations of wake width (FWHM) and amplitude (peak magnitude of the observed potential) as the probes cross the wake, as function of the solar wind ion flow Mach number, are given in Fig.~\ref{fig:alpert}. For the numerical integration of equation~\ref{eq:alpert}, the spacecraft cross section has been described as a rectangle 1x3~m in size and the probe moves across the wake 44 meters away from the centre of the spacecraft. Three different angles of the solar wind flow direction to the satellite spin plane (wake elevation angle) have been considered in Fig.~\ref{fig:alpert}. Until May 2014 the Cluster satellite spin axes were actively kept at a tilt with respect to the direction to the Sun (the Solar Aspect Angle, SAA) of typically $95^\circ\pm1^\circ$. For a solar wind flowing in the ecliptic plane, this would correspond to a wake elevation angle of $5^\circ$ in Fig.~\ref{fig:alpert}. This angle of course varies due to variations of the solar wind direction. Deviations in the solar wind direction from the average are often within 2-3$^\circ$, e.g.  \citeA{Tsyganenko2004a}, so in Fig.~\ref{fig:alpert} wake elevation angles of 3-7$^\circ$ should be most relevant. 

We note that after May 2014, the SAA remains closer to $90^\circ$ since the tilt angle is not actively controlled. This lowers spacecraft fuel consumption but interferes with high resolution EFW observations due to shadow on each probe during a short period each spin. For quasi-static (spin resolution) electric field data this can be compensated for in a similar way as for a narrow wake. To keep the wake analysis as simple as possible, this latter time period is not considered here.  

The wake width (FWHM) in Figure~\ref{fig:alpert}a is given in degrees, where $360^\circ$ defines a full spacecraft spin. The curves for the three wake elevation angles fall on top of each other, to the accuracy of the numerical evaluation. The reason for this is the essentially Gaussian shape of the wake ensured by equation~(\ref{eq:alpert}) at distances far behind (as compared to spacecraft dimensions) a spacecraft of any shape. 
The shape of a Gaussian is independent of the amplitude, which means that the observed shape of the wake will not depend on how far away from the centre of the wake a probe crosses.
Thus, we expect the measured FWHM value to be a very robust determination. 

On the other hand, the highest (absolute) value of the observed wake potential, here referred to as the wake amplitude, is a less stable measure. The wake amplitude does depend on how far away from the wake centre the probe passes during the spin, and thus on the wake elevation angle. This amplitude also depends on the electron temperature $T_e$ and the Debye length. Figure~~\ref{fig:alpert}b shows characteristic values of the wake amplitude, relevant for $KT_e = 10$~eV and short Debye lengths, so that equation~\ref{eq:boltzmann} can be used to calculate the potential. The exact numerical value can therefore not easily be compared to any single observation, but the scaling with flow angle is adequately described. For high Mach numbers and large wake elevation angles, the observed amplitude may be less than 20\% of the actual maximum voltage on the wake axis (blue curve). For small angles, the maximum amplitude increases with the Mach number, due to the decreasing ability of ions to enter the wake and fill out the density. For higher wake elevation angles the opposite effect can be seen at sufficiently fast flow ($M>10$), as the wake gets more and more narrow and in the end will only marginally reach the probe.

\begin{figure}
    \centering
    \includegraphics[width=\columnwidth]{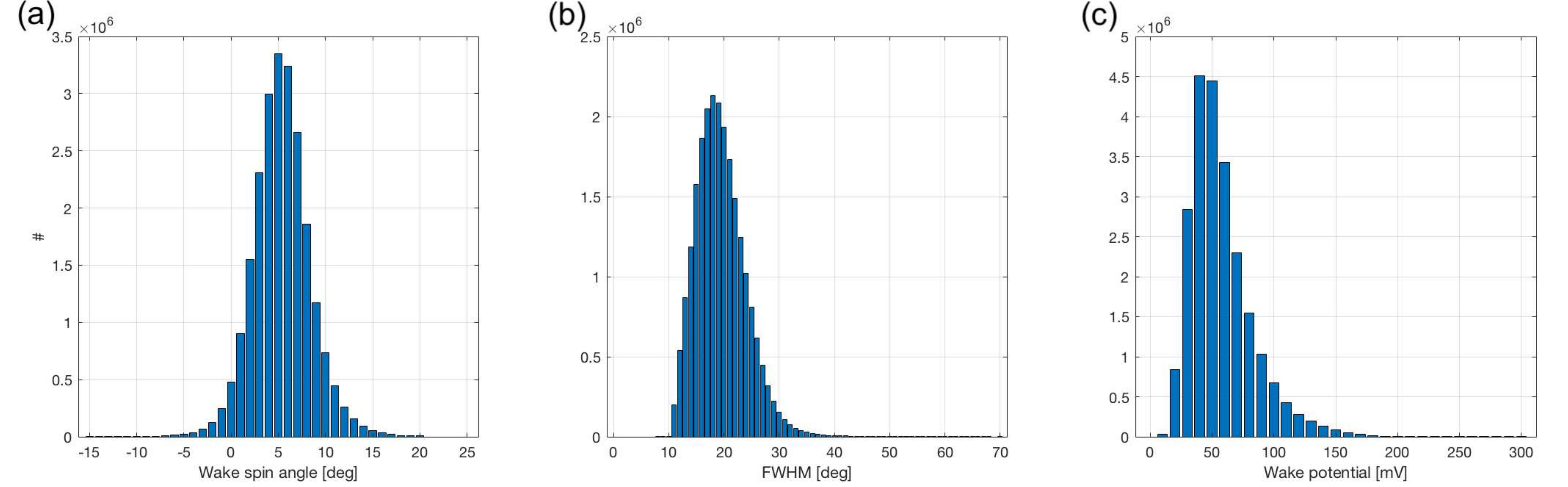}
    \caption{Solar wind wake characteristics observed by one probe pair (1-2) of the EFW instrument on Cluster~4 during three months (Jan~15 to Apr~15) of each of the years 2006-2014, in total about 22.9$ \times 10^{6}$ data points. (a) Spin phase of the wake centre (wake spin angle), with zero corresponding to the antisunward direction. (b) Full width (in degrees) at half minimum of the wake voltage signal. (c)~Wake amplitude, i.e.\ the maximum of the observed probe potential (as compared to the value outside the wake).}
    \label{fig:stat_all}
\end{figure}

To compare with observations, statistics from solar wind wake data from one probe pair (1-2) on Cluster spacecraft C4 are shown in Figure~\ref{fig:stat_all}. This figure includes 22.9$ \times 10^{6}$ identified wake signatures, each corresponding to one 4-second spacecraft spin. Observations are from 2006-2014, January 15 to April 15 each year, corresponding to the times when the orbit perigee is on the dayside and the spacecraft spend significant time in the solar wind. Data are sampled at 25~samples/s (normal mode) and sometimes 450~samples/s (burst mode), corresponding to a spin angular resolution of $3.6^\circ$ and $0.2^\circ$, respectively.
 Panel~(a) shows the wake spin angle, with zero defined as radially away from the Sun. If the solar wind flow was always radial in an inertial frame, the tangent of this angle would be the ratio of the spacecraft tangential velocity with respect to the Sun (including the orbital speed of the Earth, which dominates over the spacecraft speed around Earth) and the solar wind flow speed. The histogram could then be re-scaled to provide solar wind flow speed statistics. However, as the solar wind tangential speed is rarely zero even in a sun-fixed inertial frame, additional information on this speed must be provided to find the solar wind radial speed at any given moment. Nevertheless, by assuming that the tangential solar wind velocity (in a solar inertial frame) has a symmetrical distribution with average value of zero, we may still use Figure~\ref{fig:stat_all}a to find the mean solar wind speed for this data set. The median value of $5.0^\circ$ (with a range of 4.0 to 6.0 for the individual years) combined with the Earth's average orbital speed of 30~km/s then yields a typical solar wind radial speed of $\sim$340~km/s. In this case, this is only an order of magnitude estimate showing that the method is reasonable. The estimate of the solar wind direction deviation is reliable, but in the normal telemetry mode the typical deviation is only slightly larger than the angular resolution. We note that we have not used any selection criteria other than data quality, e.g., concerning fast and slow solar wind. In section \ref{sec:ColdIons} we use a similar technique to determine the drift velocity of ions in the polar lobes, but based on individual spacecraft spins with a well determined wake direction and using another technique to determine the perpendicular velocity.  

Panel (b) in Figure~\ref{fig:stat_all} shows the distribution of wake widths, defined by the observed FWHM, which as discussed above is expected to be a very robust observable. The median of $19^\circ$ ($15^\circ$ to $20^\circ$) can be compared to the theoretical prediction in Figure~\ref{fig:alpert}a, where it can be seen to correspond to a Mach number of about 5. For the solar wind speed of 340~km/s corresponding to the peak in solar wind direction discussed above, this yields an ion temperature of  about 20~eV, again a reasonable order of magnitude estimate for the solar wind. With the observed wake width, we can return to the quasi-neutrality assumption we introduced when using equation~(\ref{eq:boltzmann}) to estimate a theoretical value of the wake potential.
For the 44~m-long wire booms of EFW, a FWHM value of $19^\circ$ corresponds to transverse width of about 15~m across the wake. At the spacecraft location, the width of the wake is set by the spacecraft body. At 44~m from the spacecraft, the ion random thermal motion has moved ions from outside the low density region into a wider but less depleted wake. 

A wake width of 15~m is similar to the Debye length in a typical solar wind plasma with density 5~cm$^{-3}$ and electron temperature 10~eV (Table ~\ref{tab:Table_1}). For typical parameters, the Debye length is short enough for the quasi-neutrality assumption to be reasonable, and Figure~\ref{fig:alpert}b will give an order of magnitude estimate of the wake amplitude. 

Figure~\ref{fig:stat_all}c displays the maximum potential found by the probes when crossing the wake. The observations have a median of 52 mV (42-69 mV), with respect to the ambient surrounding plasma. To compare with Figure~\ref{fig:alpert}b we consider M=5-10, (consistent with typical ion temperatures and solar wind velocities), for the assumed typical electron temperature of 10~eV, and wake elevation angles of 3-7$^\circ$. This gives amplitudes of 10-30 mV, and reasonable agreement between our simple model and observations.

Our analytical model as well as particle-in-cell simulations \cite{Miyake2016}, indicate that the narrow solar wind wake extends well beyond the 44 meter EFW wire booms. 
Using this simple model, many properties of solar wind wakes can be estimated. This can be used as a tool, both to investigate the solar wind and to understand the effects on in situ observation. 
As solar wind parameters usually can be obtained by ion spectrometers, there has been little reason to develop the wake model described above to provide e.g.\ solar wind direction and ion temperature estimates. 
Rather, we use this wake model, and comparisons with particle data, to verify our understanding of basic wake physics. As we will see in next Section, there are other situations when the wake signature may give the only practical means to observe an otherwise hidden ion population.

\subsection{Wake in the polar lobes (enhanced wake)} \label{sec:Polarwake}

At high altitudes in the polar lobes the density is even lower than in the solar wind \cite{Haaland2017}. 
In this low density plasma, spacecraft charging is often high (tens of volts) since photoelectrons emitted from the satellite dominate its current balance \cite{Pedersen1995a}. The drift energy of ions originating in the ionosphere (a few eV) is often lower than the equivalent spacecraft charging, and the drift of the cold ions is often supersonic (Table \ref{tab:Table_1}), hence

\begin{equation}
     KT_{i} < m {v_{i}}^{2}/2 < eV_{SC} .
	\label{eq:Enhanced-Wake}
\end{equation}
 
 Thus, the ions are not deflected by the physical spacecraft structure but rather by a much larger electrostatic structure. This will cause an enhanced wake, Fig.~\ref{fig:figure-WakeSketchCombined}d. Also, ions will not reach the spacecraft and can not be directly detected. Some first studies of an enhanced wake behind a positively charged spacecraft are presented by \citeA{Pedersen1984} and \citeA{Bauer1983}. 
 
\begin{figure*}
\begin{center}
    \includegraphics[width=0.75\columnwidth]{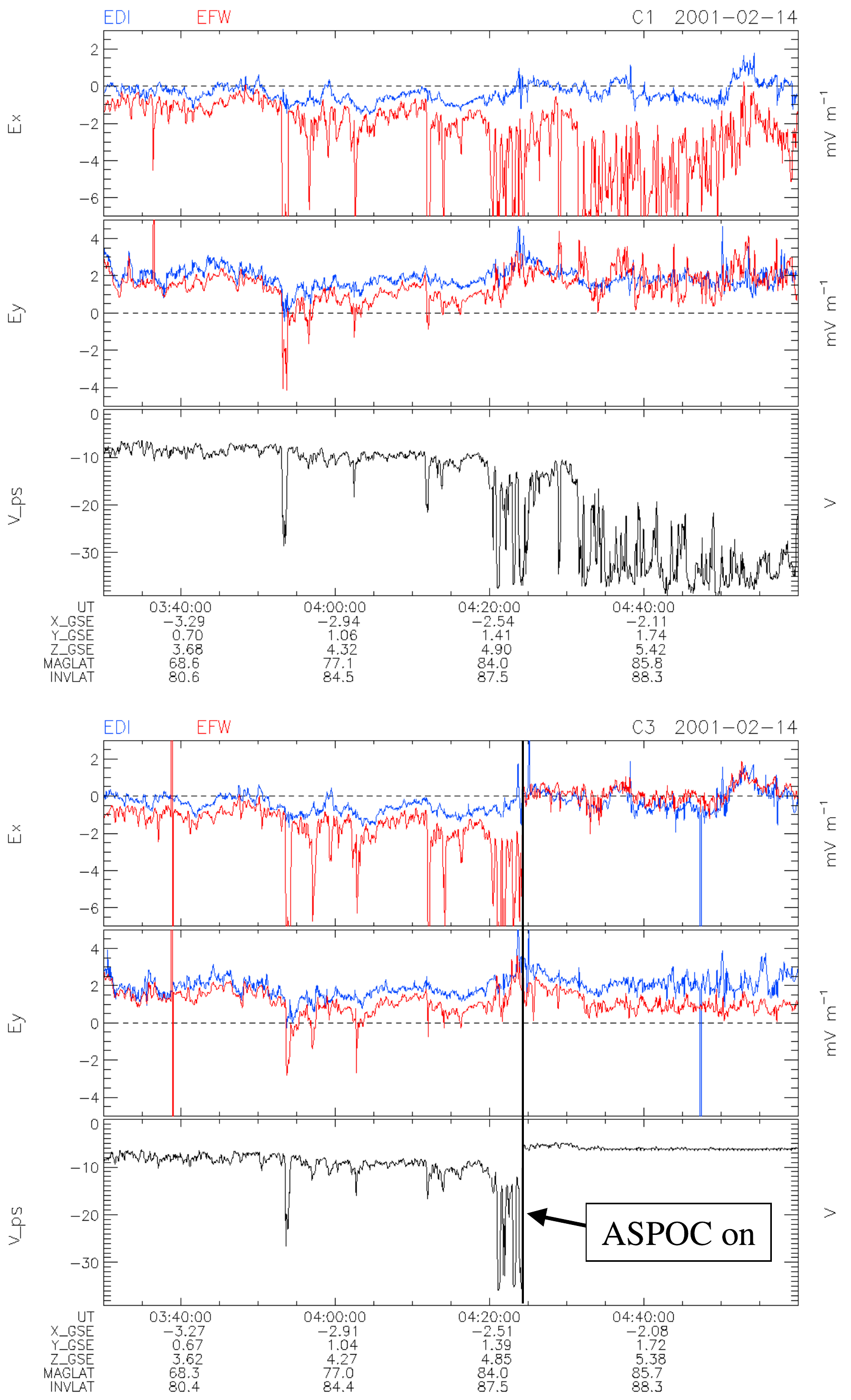}
    \end{center}
    \caption{Effects of enhanced wakes in the polar lobes. Cluster EFW (double-probe, red line) and EDI (electron drift, blue line) instrument electric field observations in the satellite spin plane, $E_{x}$ and $E_{y}$  (close to GSE x- and y-components) on spacecraft C1 and C3. The probe-to-spacecraft potential $V_{ps}$ is used to indicate the density (low $V_{ps}$ corresponds to low density and high positive spacecraft potential). During the second part of the time interval high spacecraft charging together with supersonic cold ions cause a significant local wake electric field observed by EFW. When ASPOC is turned on onboard C3 spacecraft charging and the wake are reduced, and the local wake electric field is much reduced. From \citeA{Eriksson2006}}
    \label{fig:figure-PolarCapWake_1}
\end{figure*}

 With supersonic positive ions but subsonic electrons the wake will be negatively charged. This is similar to the solar wind, but this is an enhanced wake with much larger transverse dimensions. The local wake electric field will dominate observations by a wire boom instrument, and the geophysical field can not routinely be recovered. The wake electric field can be obvious over large regions in the polar lobes. Figure~\ref{fig:figure-PolarCapWake_1} shows data from the EFW double-probe instrument (red line) and the EDI electron drift instrument (blue line) on two Cluster spacecraft (C1 and C3) \cite{Eriksson2006}. During the first part of this 1.5 hour interval the two instruments agree reasonably well most of the time. 
 The EFW probe-to-plasma potential $V_{ps}$ shown for both spacecraft is essentially the negative of the spacecraft potential $V_{SC}$ and hence indicates density variations. For conversion of $V_{ps}$ to density, see \citeA{Lybekk2012}. After 04:20 UT, $V_{ps}$ and hence the density decreases, and the spacecraft potential increases on C1. At the same time, the EFW and EDI electric fields start to clearly deviate on C1. 
 
 The large positive potential $V_{SC}$ can cause an enhanced wake when outflowing cold ions are present, relation (\ref{eq:Enhanced-Wake}). The data in Fig.~\ref{fig:figure-PolarCapWake_1} are consistent with a local (order 100 m) wake electric field observed by EFW, while the EDI observations are only marginally affected. Note that both instruments are making good observations, but one is of a local electric field dominated by an artificial field caused by the presence of a charged spacecraft, while the other is an observation over a larger region of a mainly undisturbed geophysical electric field. On C2 the ASPOC instrument is turned on at about 04:20 UT. The spacecraft potential is immediately reduced, as intended. The EFW and EDI observations become similar, further confirming the scenario of an enhanced wake which is much reduced when the spacecraft charging is reduced. A spacecraft potential of about +7 V remains, possibly causing some of the remaining difference between the EFW and EDI observations.  
 
\begin{figure*}
	\includegraphics[width=80 mm]{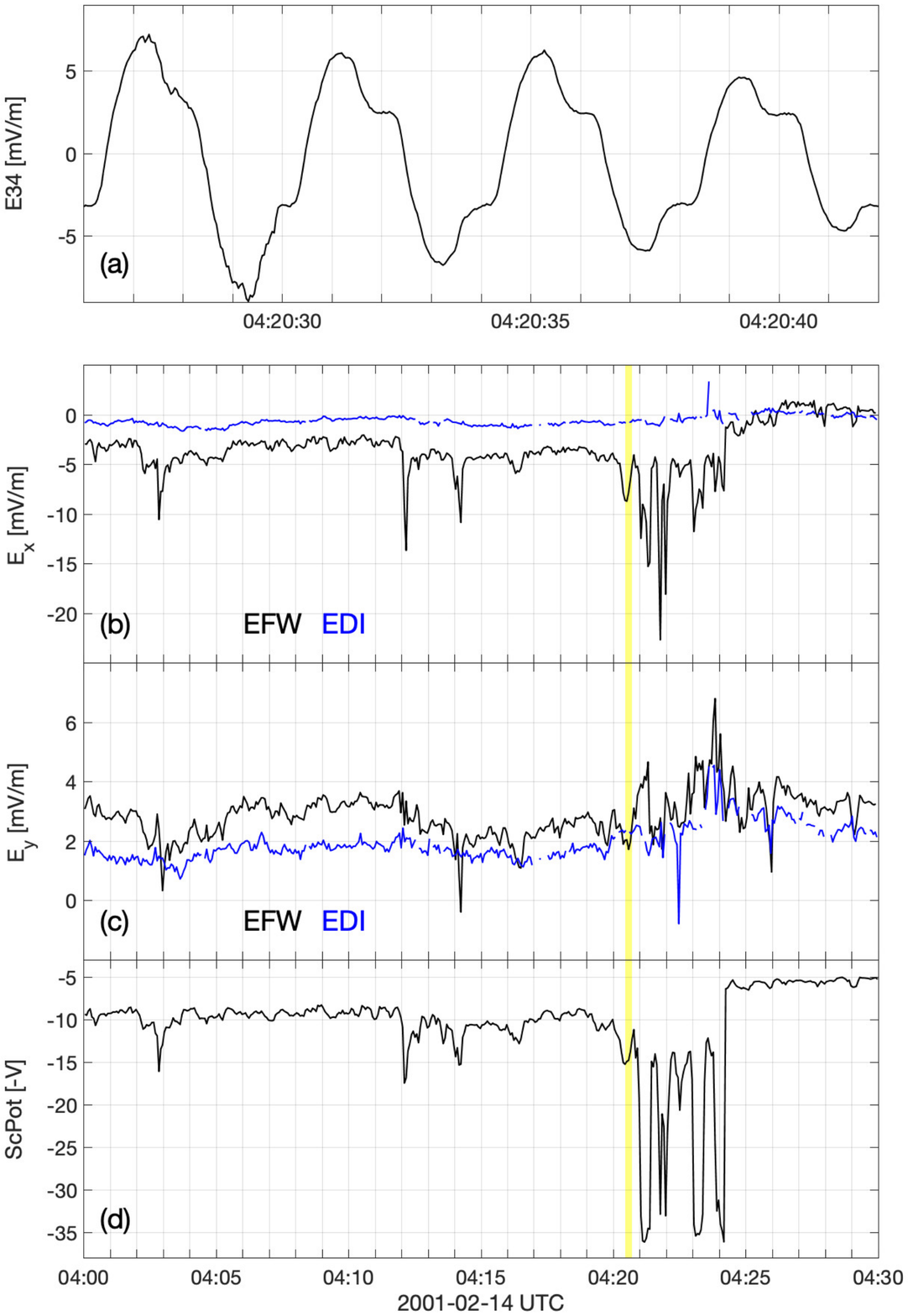}

    \caption{Effects of enhanced wakes in the polar lobes, detailed view of part of the event in Fig.~\ref{fig:figure-PolarCapWake_1} for Cluster spacecraft C3. Panel (a) shows four 4-second spins of one EFW probe-pair, from the region indicated in yellow in the lower panels. The non-sinusoidal signal indicates an intermediate size of the enhanced wake, a large wake would essentially enclose also the booms and the signal would be a sine-wave. An enhanced wake gives a large local electric field which can be used to investigate supersonic cold ions. 
    Panels (b) and (c) show EFW and EDI instrument observations of two electric field components in the satellite spin plane. Panel (d) shows the negative of the estimated spacecraft potential. ASPOC is turned on at 04:24 UT to reduce the spacecraft potential. Reducing spacecraft charging and hence the wakes makes it possible to observe the geophysical electric field with a double-probe instrument. From \citeA{Khotyaintsev2010}}.
    \label{fig:figure-PolarCapWake_2}
\end{figure*}

 Figure~\ref{fig:figure-PolarCapWake_2} shows 30 minutes of data from C3.  When ASPOC is on, the difference between EFW and EDI is much reduced. In addition, four 4-second spacecraft spins are shown from one probe-pair, when ASPOC is off. With an amplitude of a few mV/m the signal is often non-sinusoidal, as in the top panel of Figure~\ref{fig:figure-PolarCapWake_2}. For higher positive spacecraft potential (tens of volts) the signal can be sinusoidal and hard to distinguish from a geophysical quasi-static electric field. 
 
 In cases of a strongly charged spacecraft (in practise, very low density) the charged booms will give a significant contribution to the size of the extended wake. The electrostatic structure scattering cold ion can in many cases not be approximated by a sphere centered at the spacecraft and the sketch in Fig.~\ref{fig:figure-WakeSketchCombined}d is then oversimplified. However, since the ions do not reach the spacecraft the details of the scattering potential is often not of any practical importance.  
 For the case of intermediate spacecraft charging, when the ions can just marginally not reach the spacecraft (the spacecraft body has the main influence) or can indeed marginally reach the spacecraft (but effects of the charged booms must be taken into account) see section \ref{sec:MMSWwake} below. 

It is sometimes difficult to discern between local electric fields due to enhanced wakes and geophysical electric fields, and interpretation of data from double-probe instruments should be performed with caution, in particular in regions with possible cold ion drifts. For routine archiving purposes of Cluster EFW data, an algorithm is using a combination of parameters including spacecraft potential, magnetic field direction and different electric field components. When the magnetic field is close to the Cluster spin plane, the algorithm searches for indications of a large local parallel electric field. (A large geophysical electric field parallel to the magnetic field would give high-energy particles, which are not observed.) For other magnetic field directions, different perpendicular components of the electric field are compared (assuming zero parallel electric field, since observations are obtained only in the spin plane.) Higher ratios indicate a higher probability of an enhanced wake. For more focused investigations, when EDI data are available, significant differences between EFW and EDI observations can be used as an indication of a wake. Sometimes a combination of wake and geophysical electric fields, observed by EFW and EDI, can be used for scientific investigations, see section \ref{sec:ColdIons} on ionospheric outflow below.  

For a narrow solar wind wake (section~(\ref{sec:SWwake})), the wake electric field is observed by EFW during a small part of the spacecraft spin. Here the wake signature can removed, and the geophysical electric field can be obtained in many directions (Fig.~\ref{fig:figure-SolarWindWake}). For an enhanced wake in the lobes, the electric field observed by EFW is again a sum of a wake field and a geophysical field. But here the wake field is observed during the whole spacecraft spin, Fig.~\ref{fig:figure-PolarCapWake_2} \cite{Khotyaintsev2010}. \citeA{Engwall2006b} showed examples indicating that it is in principle possible to obtain the geophysical electric field from the EFW instrument also for an enhanced wake, by considering the Fourier spectrum of the observed signal. This requires that the spin-period signal from one probe-pair is not a sinusoidal (some signal from the geophysical field can be detected). The spin tone harmonics in this spectrum are due only to the wake, whose direction thereby can be determined and the wake removed. 
This method is complicated to use, partly due to the the so-called sunward offset \cite{Cully2003, Khotyaintsev2010, Khotyaintsev2014} but can in principle be attempted on an event basis. 
Our observations, and also simulations \cite{Engwall2006, Eriksson2010b, Miyake2016}, indicate that the enhanced polar lobe wake extends well beyond the 44 meter EFW wire booms. There is no attempt to routinely obtain the geophysical electric field but this situation is used for statistical investigations of the flux of cold ions, see section \ref{sec:ColdIons}.


\subsection{Intermediate parameters} \label{sec:MMSWwake}

In an intermediate parameter range, supersonic cold ions can marginally reach the charged spacecraft but are significantly affected by both the charged spacecraft and the charged wire booms of an electric field instrument. In this case

\begin{equation}
m {v_{i}}^{2}/2 \gtrsim eV_{SC} , \; \; \; \; \;  m {v_{i}}^{2}/2 >  KT_{i} .
\label{eq:MMS-Wake}
\end{equation}

\begin{figure*}
	\includegraphics[width=80 mm]{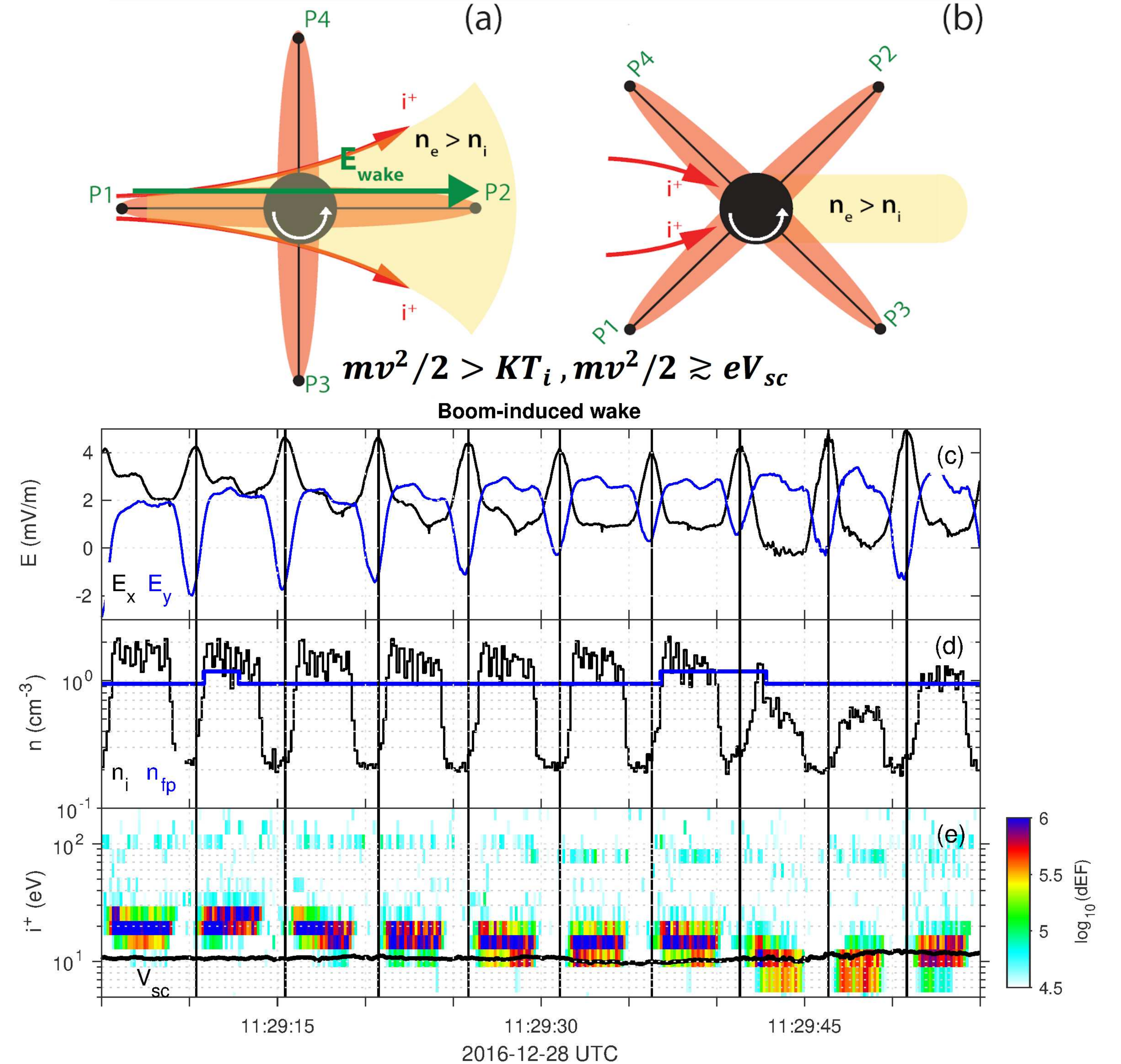}
    \caption{Sketch of one MMS spacecraft with wire booms in a flow of ions (see also Fig.~\ref{fig:figure-WakeSketchCombined}). Panels (a)  and (b): Sketch of two phases of the 20 s spacecraft spin. Positive potential around the wire booms is indicated in orange, negative space charge in the wake is indicated in yellow, positive ion orbits are shown in red. (c) Two components of the electric field, (d) density obtained from ion ($n{_i}$) data and from the plasma frequency ($n_{lp}$) (e) ion flux and the spacecraft potential, see \citeA{Toledo-Redondo2019}. For a supersonic ion flow with drift velocity similar to the equivalent spacecraft charging, the charged wire booms have large influence on the ion orbits and cause a periodic behaviour of observed particles and electric fields. 
    }
    \label{fig:figure-MMS_Wake_v2}
\end{figure*}

Figure~\ref{fig:figure-MMS_Wake_v2} illustrates how the electric field instrument wire booms on MMS are important for the ions trajectories, in this case just inside the magnetopause \cite{Toledo-Redondo2019}. The upper part of the figure shows sketches of a changing situation as the spacecraft spins: Ions are deflected by the electric field of charged booms and can not reach particle detectors on the spacecraft, or the ions are focused into on-board detectors, see also the simulations by \citeA{Miyake2013}. The wake behind the spacecraft changes as a function of the spin phase, and the electrostatic potential structure cannot be approximated as spherical. The three lower panels show MMS observations of this effect. Fig.~\ref{fig:figure-MMS_Wake_v2}c shows the electric field in the spin plane. Every $\sim$5 s, i.e., a quarter of the MMS spin period, the double probes measure a non-geophysical wake electric field (marked with vertical black lines), while the electric field measured between the electric field spikes is a geophysical field which is supported by a good agreement between the measured \textbf{E} and -{\bf v}$\times${\bf B} (not shown). 
Fig.~\ref{fig:figure-MMS_Wake_v2}d shows the ion density, measured using an ion detector (black), and inferred from the plasma frequency (blue). An artificial dropout in plasma density is measured by the ion detector when the wire booms are aligned to the cold ion flow which is then deflected, as illustrated in Fig.~\ref{fig:figure-MMS_Wake_v2}a. Density enhancements are also observed by the detector between the vertical black lines, which are consistent with Fig.~\ref{fig:figure-MMS_Wake_v2}b, although no independent validation of the calibration of the low-energy channels of the ion instrument has been performed for this time period. 
Fig.~\ref{fig:figure-MMS_Wake_v2}e shows the omnidirectional spectrogram  recorded by the ion instrument and the spacecraft potential (black line). The cold proton beam has drift energies of about 2 times the equivalent spacecraft potential, and the repetitive detection gaps every quarter of spin can be clearly observed. The light blue signature at $\sim$100 eV corresponds to cold He$^+$, and detection gaps near the vertical black lines can also be observed, despite their drift energy is about 8 times larger than the spacecraft potential. This can be attributed to deflection of the ions by the electric fields pointing outward from the changed wire booms. See also \citeA{Barrie2019} for an additional discussion on particle orbits near the charged MMS satellites.   

Care must be taken not to confuse periodic behaviour of electric field and particle data (Fig.~\ref{fig:figure-MMS_Wake_v2}) with natural wave phenomena. A clear warning sign is a steady periodicity at a multiple of the satellite spin frequency. Also, when the spacecraft charging is similar to the equivalent ion drift energy (at the magnetopause, often $\mathbf{E \times B}$ drift) a spherically symmetric potential structure around the spacecraft body can not be used to correct particle observations \cite{Toledo-Redondo2019}. The example in Fig.~\ref{fig:figure-MMS_Wake_v2} is unusually clear but particle moments may be affected by asymmetric charging also when periodic effects are not so obvious.  

\section{The enhanced wake as a tool to detect cold ions} \label{sec:ColdIons}

It has been suggested for decades that cold ions from the high-latitude ionosphere can dominate the density and outflow in the high-altitude magnetospheric tail lobes \cite{Chappell1980, Moore1984, Olsen1985, Chappell2015}. These positive ions often have a drift energy of one or a few eV, and even lower thermal energy, and hence can not reach a spacecraft charged positively to tens of volts. Such a supersonic outflowing "polar wind" was predicted by \citeA{Axford1968} and \citeA{Banks1968}. There are several studies of outflowing ions in the polar regions at altitudes up to a few Earth radii \cite{Cully2003, Abe2004, Huddleston2005, Peterson2006, Peterson2008, Nilsson2013},  see reviews by \citeA{YauAndre1997}, \citeA{Yau2007}, \citeA{Moore2007}, \citeA{AndreCully2012}, \citeA{Welling2015}, 
\citeA{Yamauchi2019}, \citeA{Delzanno2021}, \citeA{Yau2021}, \citeA{Andre2021} and \citeA{Toledo-Redondo2021}. However, at higher altitudes many ions can not reach a positively charged spacecraft. On the Polar spacecraft the charging could during some periods be artificially reduced down to a few volts positive by emitting a plasma cloud but still a significant fraction of the cold outflowing ions could be missed \cite{Moore1997, Su1998, Engwall2009b}. An alternative method based on Cluster observations does not depend on the ions reaching the spacecraft, but is rather using the enhanced wake induced by the drifting cold ions to estimate the flux of these ions \cite{Engwall2009a, Engwall2009b}. While the enhanced wakes make observations of the geophysical electric field with a double-probe instrument complicated and often impossible, these wakes make it possible to detect a previously hidden cold ion population.

The wake-method to estimate the cold ion drift velocity is based on the local electric field (observed by the EFW double-probe instrument) combined with the large-scale geophysical electric field (observed by the EDI instrument). The wake electric field is obtained as the difference between the local and the geophysical electric fields. In the lobes the ions can be treated as unmagnetized on the wake length scale (Table \ref{tab:Table_1}) and the direction of the wake electric field gives the ion drift direction. The ion drift perpendicular to the ambient magnetic field is given by the geophysical electric field (EDI) and magnetic field observations from the Fluxgate Magnetometer (FGM) \cite{Balogh2001}. Since the perpendicular velocity component and the direction of the flow are known, the parallel component can be inferred. 

The density can be estimated by calibrating observations of the spacecraft potential obtained by the Cluster EFW instrument \cite{Pedersen2008, Svenes2008, Lybekk2012, Haaland2012b}. The potential induced by the wake is small, tens of millivolts (Fig.~\ref{fig:stat_all}), compared to the spacecraft potential of tens of volts (Fig.~\ref{fig:figure-PolarCapWake_1}), and has negligible effect on this estimate.
The density and the outflow velocity give the ion flux. 

The method of using an enhanced wake to estimate the flux of cold supersonic ions has been validated in multiple ways. In situ comparison of two Cluster spacecraft in the same polar lobe region give similar results, with one spacecraft using the wake method and the other using particle detectors and artificial reduction of the spacecraft potential \cite{Engwall2006c}. The method has also been studied using simulations \cite{Engwall2006}. Our investigations in the solar wind shows that our understanding of wake physics using the Cluster instruments is correct (section \ref{sec:SWwake}). Also, the order of magnitude of the outflow of cold ionospheric ions is the same for the Cluster wake method and for observations obtained with particle detectors on other spacecraft, at much lower altitude where spacecraft charging is less of a problem, or by some studies using observations when artificially reducing this charging (see recent summaries by \citeA{Andre2021} and \citeA{Toledo-Redondo2021}). Even when reducing spacecraft charging typically a charge of a few volts positive remains and such observations can not replace the wake method in the low-density polar lobes.  

In summary, the presence of a supersonic flow of low-energy ions can be inferred by detecting a wake electric field, obtained as large enough difference between the quasi-static electric fields observed by the EFW (total electric field) and EDI (geophysical electric field) instruments. To estimate the parallel drift velocity, observations of the perpendicular $\mathbf{E \times B}$ drift velocity from the geophysical quasi-static electric field (EDI) and the geophysical magnetic field (FGM) are needed, together with the direction of the wake electric field. The ion flux can then be estimated from the drift velocity and the density. Details concerning the data analysis and error estimates are given by \citeA{Engwall2009b} and in Appendix A of \citeA{Andre2015a}.

\begin{figure*}
	\includegraphics[width=80 mm]{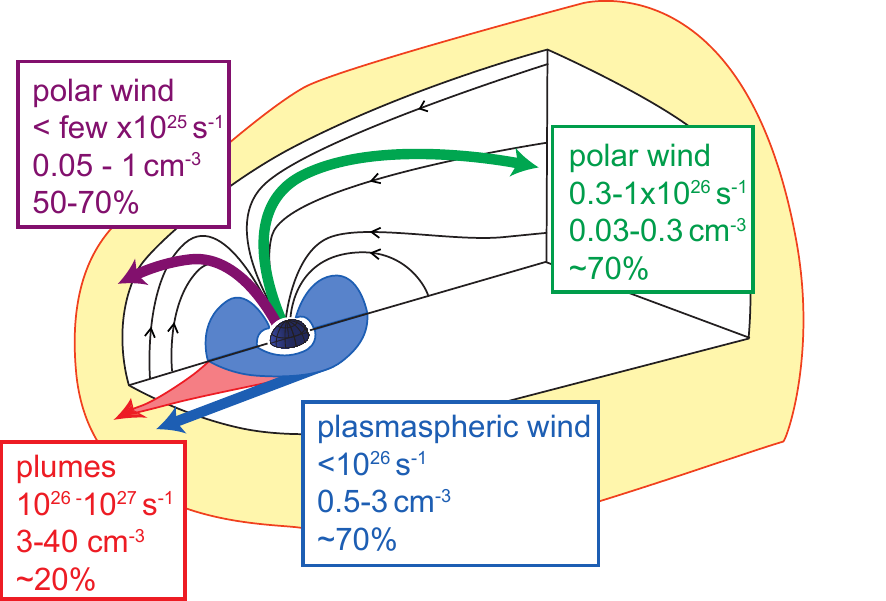}
    \caption{Overview of cold (eV) ion outflow. Typical outflow rates and densities are given together with the approximate fraction of time cold ions dominate the number density. For high latitudes, this fraction is estimated from observations of enhanced spacecraft wakes indicating cold ions with supersonic drift. For the magnetopause, a combination of methods is used. Cold ions often dominate the density of the magnetosphere. The drift paths are not obtained from local observations and are discussed in several studies, see text for references. (Figure from \citeA{AndreCully2012}). 
    }
    \label{fig:figure-Overview_GRL_2012}
\end{figure*}

One ion flux estimate can be obtained for each 4-second Cluster spacecraft spin \cite{Engwall2009a, Engwall2009b}. Even when applying rather strict limits to minimize errors, 320,000 data points (satellite spins) can be used from early 2001 to 2010 (from the peak of solar cycle 23 to beyond the minimum of solar cycle 24) \cite{Andre2015a}. The low-energy ions usually dominate the density and the outward flux in the geomagnetic tail lobes during all parts of the solar cycle. The wake method does not determine the mass of the outflowing ions, but most are believed to be low-mass H$^{+}$. Heavier ions such as O$^{+}$ would have higher energy than lighter ions for a given drift velocity. These ions would be easier to detect onboard a charged spacecracft and would then not contribute to an enhanced wake. Also, observations at lower altitudes with less spacecraft charging, and also observations using artificial reduction of the spacecraft charging, indicates that most ions are H$^{+}$ \cite{Su1998}.
The global outflow is of the order of $10^{26}$ ions/s and often dominates over the outflow at higher energies \cite{Engwall2009a, AndreCully2012, Andre2015a}. Depending on overall geophysical conditions the ions may not immediately leave the magnetosphere \cite{Haaland2012b} but are likely to eventually be lost to the solar wind \cite{Andre2015a, Andre2021}. This outflow is a significant part of the total mass outflow from Earth \cite{Andre2015b}.
Figure~\ref{fig:figure-Overview_GRL_2012} shows an overview of low-energy ion outflow. The Cluster wake-method to detect cold ions has been a major method to obtain this overall picture.
 

\section{Summary}

Wakes in collisionless plasmas are common, both behind spacecraft and other obstacles. 
Behind spacecraft, wakes caused by positive supersonic ions are a well known problem affecting in situ observations, including electric field observations. 

We review the impact of wakes on observations by the Electric Field and Wave instrument on the Cluster satellites. Due to the polar orbit, Cluster can investigate both the solar wind and the low-density polar lobes. In the solar wind the effects of the wake are minor, easy to detect, and can be compensated for in a reasonable way. In the polar lobes the effects of the wake are major, due to an enhanced effect caused by a very positively charged spacecraft, and makes observations of the geophysical electric field complicated or impossible, at least close to the satellite. In this situation detection of the wake using the EFW double-probe instrument can be used to detect supersonic drifting cold ions. Together with other instruments also the ion flux can be estimated. These enhanced wakes are clearly different not only from the wakes in the solar wind but also from spacecraft wakes in low earth orbit and from wakes behind natural objects such as charged dust and the Moon. 

The charging of the long wire booms of a double-probe instrument contributes to the electrostatic structure scattering drifting cold ions. For a very charged spacecraft, typical for the polar lobes, the details of this electrostatic structure can often be ignored when interpreting observations.  
For an intermediate range of parameters, when the drift energy of the cold ions is similar to the equivalent spacecraft charging, also the charging of the wire booms must be considered in detail when interpreting data. 

Some common phenomena related to the Cluster EFW double-probe instrument are not discussed in detail here. One example is the spurious electric fields in the plasmasphere. Fields that are not geophysical of the order 1-2 mV/m, mainly in the sunward direction, are detected by an empirical algorithm, \cite{Puhl-Quinn2008, Khotyaintsev2014}. This spurious field seems partly related to a subsonic ion flow and the long wire booms \cite{Miyake2015}.

Plasma wakes behind spacecraft with instruments for in situ plasma observations are common. These wakes change the local plasma environment, as compared to the geophysical conditions without the spacecraft. The wakes can make some observations of geophysical parameters complicated, and sometimes impossible. With understanding of the physics causing the wakes, the local effects can in many situations be compensated for. In some situations otherwise inaccessible geophysical parameters can be estimated, using the wake caused by the presence of the spacecraft. An important example is the flux of cold positive ions in the polar lobes. This flux of the order of 10$^{26}$ ions/s constitutes a significant part of the mass outflow from planet Earth. Often these positive ions can not reach a positively charged spacecraft. Rather, the ion flux can be obtained from the properties of the enhanced wake. 

\begin{table}[ht]
    \hskip-1.0cm\begin{tabular}{lllllll}
    \multicolumn{7}{l}{\textbf{Examples of parameters: Low Earth Orbit, upper ionosphere}}\\
    $n$ (cm$^{-3}$) & $KT_{e}$ (eV) & $KT_{i}$ (eV) & $B$ (nT) & ion & $v_{i}$ (km/s)  & $V_{SC}$ (V)\\
    \hline
     2$\cdot$10$^{5}$    & 0.3 & 0.2 & 50000 & H$^{+}$ or O$^{+}$ & 8 & -1 \\
     \hline
     \hline
      ${\lambda}_{D}$ (m) & ${\rho}_{e}$ (m) & ${\rho}_{i}$ (m) & $v_{th}$ (km/s) & $ m {v_{i}}^{2}/2 $ (eV) & \\
      \hline
      0.01 &                 0.024 &        0.85                 &  5.8             & 0.3 & \\
    \hline
    \hline
      \multicolumn{7}{l}{Useful relations}\\
      \hline
       ${\lambda}_{D} < L_{SC}$ & $\;  {\rho}_{e} <  L_{SC}$  & $\;   {\rho}_{i} \approx L_{SC}$  & \\
      \hline
      \hline
      \multicolumn{7}{l}{\textbf{Examples of parameters: Solar wind (narrow wake)}}\\
    $n$ (cm$^{-3}$) & $KT_{e}$ (eV) & $KT_{i}$ (eV) & $B$ (nT) & ion & $v_{i}$ (km/s)  & $V_{SC}$ (V)\\
    \hline
      5   & 10 & 10 & 5 & H$^{+}$  & 400 & +5 \\
     \hline
     \hline
      ${\lambda}_{D}$ (m) & ${\rho}_{e}$ (m) & ${\rho}_{i}$ (m) & $v_{th}$ (km/s) & $ m {v_{i}}^{2}/2 $ (eV) & \\
      \hline
      10 &                 1400 &        60000                &  41             & 830 & \\
    \hline
    \hline
      \multicolumn{7}{l}{Useful relations}\\
      \hline
      $KT_{i} < {m v_{i}}^{2}/2$  & $\; m {v_{i}}^{2}/2 >> eV_{SC}$ & $\; {\lambda}_{D} > L_{SC}$ & $\; {\rho}_{e} >> L_{boom}$ & $\; {\rho}_{i} >> L_{boom}$  \\
      \hline
      \hline
       \multicolumn{7}{l}{\textbf{Examples of parameters: Polar lobes (enhanced wake)}}\\
    $n$ (cm$^{-3}$) & $KT_{e}$ (eV) & $KT_{i}$ (eV) & $B$ (nT) & ion & $v_{i}$ (km/s)  & $V_{SC}$ (V)\\
    \hline
      0.1   & 2 & 1 & 20 & H$^{+}$  & 30 & +40 \\
     \hline
     \hline
      ${\lambda}_{D}$ (m) & ${\rho}_{e}$ (m) & ${\rho}_{i}$ (m) & $v_{th}$ (km/s) & $ m {v_{i}}^{2}/2$ (eV) & \\
      \hline
      30 &                 160 &        4800                &  13             & 5 & \\
    \hline
    \hline
      \multicolumn{7}{l}{Useful relations}\\
      \hline
      $ KT_{i} < m {v_{i}}^{2}/2 << eV_{SC} $& $\;  {\lambda}_{D} >> L_{SC}$ & $\; {\rho}_{e} >>  L_{SC}$ &  $\; {\rho}_{e} > L_{boom}$ & $\; {\rho}_{i} >> L_{boom}$ \\ 
      \hline
      \hline
        \multicolumn{7}{l}{\textbf{Examples of parameters: Spacecraft dimensions}}\\
\hline
   Spacecraft body  &  Wire booms \\ 
   $L_{SC} \approx 2$ m &  $L_{boom} \approx 100$ m & & & & & \\
   \hline
   \end{tabular}
    \caption{Examples of parameters for LEO in the upper ionosphere, solar wind and polar lobes. Here $n$, $KT_{e}$, $KT_{i}$, $B$, $v_{i}$ and $V_{SC}$ are the density, electron and ion thermal energies, geomagnetic field, ion drift velocity and spacecraft potential. From this we derive ${\lambda}_{D}$, ${\rho}_{e}$, ${\rho}_{i}$, $v_{th}$ $ m {v_{i}}^{2}/2 $, the Debye length, electron and ion gyroradii, thermal ion velocity and ion drift energy. 
    Typical length scales of a spacecraft main body and wire booms are also given, $L_{SC}$ and $L_{boom}$. In LEO, the drift velocity is taken to be the velocity of an orbiting spacecraft, while a sounding rocket moves much slower, and derived parameters are given for H$^{+}$. In the solar wind, the drift velocity is an example of a solar wind velocity, and in the polar lobes a typical outflow velocity of ionospheric ions is given. For an overview of near-Earth plasma parameters see textbooks, e. g. \citeA{KivelsonRussell1995}. 
    Relevant parameters for LEO (sounding rockets and the International Space Station) are given by \citeA{Paulsson2019} and \citeA{Reddell2006}, from the solar wind by  \citeA{Eriksson2006, Eriksson2007b}, and from the polar lobes by \citeA{Engwall2009b}, \citeA{Andre2015a} and \citeA{Haaland2017}. Sketches of corresponding wakes are given in Figure \ref{fig:figure-WakeSketchCombined}.
    }
    \label{tab:Table_1}
\end{table}

\acknowledgments
MA is supported by Swedish National Space Agency contract 2020-00058. We are grateful for support from the Cluster and MMS instrument teams. Cluster data are available from the Cluster Science Archive https://www.cosmos.esa.int/web/csa and MMS data are available from https://lasp.colorado.edu/mms/sdc/public/.
We acknowledge support from the ISSI international team Cold plasma of ionospheric origin at the Earth’s magnetosphere. 


%
%

\bibliography{WakePaper}

%
%
%
%
%

\end{document}